\documentstyle[11pt]{article}
\newfont{\bbbold}{msbm10 scaled \magstep1}

\def\cE{\cal E}
\def\cF{{\cal F}}

\def\cL{\cal L}

\newfont{\goth}{eufm10 scaled \magstep1}

\def\a{\alpha}
\def\b{\beta}
\def\c{\gamma}\def\C{\Gamma}
\def\d{\delta}
\def\e{\epsilon}

\def\h{\eta}
\def\k{\kappa}
\def\L{\Lambda}
\def\m{\mu}

\def\P{\Pi}

\def\th{\theta}

\def\be{\begin{equation}}\def\ee{\end{equation}}
\def\bea{\begin{eqnarray}}\def\eea{\end{eqnarray}}
\def\ba{\begin{array}}\def\ea{\end{array}}

\def\o{\omega}\def\O{\Omega}
\def\del{\partial}
\def\ua{\underline{\alpha}}
\def\ub{\underline{\phantom{\alpha}}\!\!\!\beta}

\def\um{\underline{\mu}}

\def\una{\underline a}\def\unA{\underline A}
\def\unb{\underline b}\def\unB{\underline B}
\def\unc{\underline c}\def\unC{\underline C}
\def\und{\underline d}

\def\unm{\underline m}\def\unM{\underline M}
\def\unN{\underline N}
\def\unP{\underline P}

\def\nab{\nabla}

\def\del{\partial}

\let\la=\label

\let\bm=\bibitem
\def\um{{\underline m}}

\def\nn{\nonumber}
\def\bd{\begin{document}}
\def\ed{\end{document}}
\def\bea{\begin{eqnarray}}
\def\eea{\end{eqnarray}}
\def\ft#1#2{{\textstyle{{\scriptstyle #1}\over {\scriptstyle #2}}}}
\def\fft#1#2{{#1 \over #2}}
\newcommand{\eq}[1]{(\ref{#1})}
\def\eqs#1#2{(\ref{#1}-\ref{#2})}
\def\det{{\rm det\,}}
\def\tr{{\rm tr}}
\newcommand{\ho}[1]{$\, ^{#1}$}
\newcommand{\hoch}[1]{$\, ^{#1}$}
\newcommand{\tamphys}{\it\small Center for Theoretical Physics, 
Texas A\&M University, College Station, TX 77843, USA}
\newcommand{\newton}{\it\small Isaac Newton Institute for Mathematical
Sciences, Cambridge, UK}
\newcommand{\kings}{\it\small Department of Mathematics, King's College,
London, UK}

\newcommand{\auth}{\large P.S. Howe\hoch{1}, O. Raetzel\hoch{1} and E.
Sezgin \hoch{2\dagger} }

\begin{document}

\hfill{KCL-MTH-98-10}

\hfill{CTP TAMU-11/98}

\hfill{hep-th/9804051}

\vspace{20pt}

\begin{center}
{\Large{\bf On Brane Actions and Superembeddings}}
\vspace{30pt}

\auth

\vspace{15pt}

\begin{itemize}
\item [$^1$] \kings
\item [$^2$] \tamphys
\end{itemize}

\vspace{60pt}

{\bf Abstract}

\end{center}

Actions for branes, with or without worldsurface gauge fields, are
discussed in a unified framework. A simple algorithm is given for
constructing the component Green-Schwarz actions. Superspace actions are
also discussed. Three examples are given to illustrate the general
procedure: the membrane in $D=11$ and the D2-brane, which both have
on-shell worldsurface supermultiplets, and the membrane in $D=4$, which
has an off-shell multiplet.

{\vfill\leftline{}\vfill
\vskip	10pt
\footnoterule
{\footnotesize
\hoch{\dagger} Research supported in part by 
NSF Grant PHY-9722090 \vskip -12pt} 

\pagebreak
\setcounter{page}{1}

%%%%%%%%%%%%%%%%%%%%%%%%%%%%%%%%%%%%%%%%%%%%%%%%%%%%%%%%%%%%%%%%%%%%%%%%%%%%

\section{Introduction}

%%%%%%%%%%%%%%%%%%%%%%%%%%%%%%%%%%%%%%%%%%%%%%%%%%%%%%%%%%%%%%%%%%%%%%%%%%%%

In the superembedding approach to supersymmetric extended objects the
object under consideration is described mathematically as a
subsupermanifold (the worldsurface) of superspacetime (the target
supermanifold). This approach was initiated some time ago \cite{e1,e2}
in the context of superparticles. Superspace actions were found for
particles in $D=3,4,6$ and $10$ dimensional spacetimes
\cite{e1,e2,e3,e4,e5,e6} and then later for the heterotic string in ten
dimensions \cite{e7,e8}. Actions of the heterotic string type were
constructed for other type I branes (i.e. branes with no worldsurface
vector or tensor fields) \cite{e9,e10} but it was not clear at the time
that these actions described the right degrees of freedom. In \cite{a1}
a generalised action was proposed for type I branes which leads to the
standard Green-Schwarz equations of motion (see \cite{a2,a3} for
reviews) and this approach has recently been extended to cover $D$-branes
\cite{a4,a5,a6}.

The structure of the worldsurface supermultiplets that arise in the
superembedding formalism was clarified in \cite{hs1}; there it was
assumed that a natural embedding condition, namely that the odd tangent
bundle of the worldsurface should be a subbundle of the pull-back of the
odd tangent bundle of the target space, holds. It was found in
\cite{hs1} that three types of multiplet can arise: on-shell, off-shell
or underconstrained. In the on-shell case, there can be no superspace
actions of the heterotic string type since such actions would
necessarily involve the propagation of the Lagrange multipliers that are
used in this construction. Nevertheless, on-shell embeddings are useful
for deriving equations of motion; for example, the full equations of
motion of the $M$-theory fivebrane were first obtained this way
\cite{hs2}. In the off-shell case, by which it is meant that the
wordsurface multiplet is a recognisable off-shell multiplet, it is
possible to write down actions of the heterotic string type. The third
case that arises, and which we call underconstrained here, typically
occurs for branes with low codimension. For example, in codimension one
the basic embedding condition gives rise to an unconstrained scalar
superfield. In order to get a recognisable multiplet further constraints
must be imposed. An example of this is given by IIA D-branes where the
basic embedding condition yields an on-shell multiplet for $p=0,2,4$,
but an underconstrained one for $p=6,8$. By imposing by hand the further
constraint that there is a worldsurface vector field with the usual
modified Bianchi identity whose superspace field strength vanishes
unless all indices are bosonic one recovers on-shell multiplets
\cite{hrss}. (For $p=0,2,4$ one can show that the vector Bianchi
identity follows from the basic embedding condition.)

In this note we show that there is a simple algorithm for generating
actions for (almost) all branes starting from the superembedding
formalism. It can be used in two ways: if the multiplet is on-shell, one
can use it to find the Green-Schwarz action; if the multiplet is
off-shell one can use it either to write down a superspace action of
heterotic string type or one can construct a Green-Schwarz action which
in general will have auxiliary fields. In the underconstrained case we
shall assume that further constraints have been imposed to convert the
embedding into one of the first two types. The actions obtained this way
are Lorentz covariant and are thus not applicable to branes with
self-dual tensor multiplets, although actions involving additional
fields have been proposed for these cases \cite{ma1,ma2}. We give three
examples: the $D=11$ supermembrane, which has an on-shell scalar
multiplet, the type IIA $D2$-brane in $D=10$, which has an on-shell
vector multiplet, and the membrane in $D=4$, which is off-shell. 

The method of constructing actions proposed here is closely related both
to the superspace method used for the heterotic string and to the
generalised action principle. However, the proof that the GS action is
$\k$-symmetric is greatly simplified. In addition, our approach is
deductive in the sense that we derive the GS action from the
superembedding formalism. Thus, in the case of D-branes, rather than
starting with the Dirac-Born-Infeld (DBI) term in the action we show
that it emerges from the construction. An advantage of this approach is
that it is applicable to other type II branes which have higher rank
worldsurface antisymmetric tensor gauge fields, provided that they are not self-dual.
%%%%%%%%%%%%%%%%%%%%%%%%%%%%%%%%%%%%%%%%%%%%%%%%%%%%%%%%%%%%%%%%%%%%%%%%%%%%

\section{Superembeddings}

%%%%%%%%%%%%%%%%%%%%%%%%%%%%%%%%%%%%%%%%%%%%%%%%%%%%%%%%%%%%%%%%%%%%%%%%%%%%

We consider superembeddings $f:M\rightarrow \unM$, where the
worldsurface $M$ has (even$|$odd) dimension $(d|{1\over2}D')$ and the
target space has dimension $(D|D')$. In local coordinates $M$ is given
as $z^{\unM} (z^M)$, where $z^{\unM}=(x^{\unm},\th^{\um})$ and
$z^{M}=(x^{m},\th^{\m})$ (if no indices are used we shall distinguish
target space coordinates from worldsurface ones by underlining the
former). The embedding matrix $E_A{}^{\unA}$ is defined to be

\be
E_A{}^{\unA}= E_A{}^M\del_M z^{\unM} E_{\unM}{}^{\unA}\ ,
\label{1}
\ee

in other words, the embedding matrix is the differential of the
embedding map referred to standard bases on both spaces. Our index
conventions are as follows: latin (greek) indices are even (odd) while
capital indices run over both types; letters from the beginning of the
alphabet are used to refer to a preferred basis while letters from the
middle of the alphabet refer to a coordinate basis, the two types of
basis being related to each other by means of the vielbein matrix
$E_M{}^A$ and its inverse $E_A{}^M$; exactly the same conventions are
used for the target space and the worldsurface with the difference that
the target space indices are underlined. Primed indices are used to
denote directions normal to the worldsurface. We shall also use a
two-step notation for worldsurface spinor indices where appropriate: in
general discussions, a worldsurface spinor index such as $\a$ runs from
1 to ${1\over2}D'$, but it may often be the case that the group acting
on this index includes an internal factor as well as the spin group of
the worldsurface; in this case we replace the single index $\a$ with the
pair $\a i$ where $i$ refers to the internal symmetry group. A similar
convention is used for normal spinor indices.

The basic embedding condition is

\be
E_{\a}{}^{\una}=0\ .
\label{2}
\ee

It implies that the odd tangent space of the worldsurface is a subspace
of the odd tangent space to $\unM$ at each point in $M\subset \unM$. In
many cases, equation \eq{2} determines the equations of motion for the
brane under consideration. Moreover, it also determines the geometry
induced on the worldsurface and implies constraints on the background
geometry which arise as integrability conditions for the existence of
such superembeddings. For the cases where the worldsurface multiplet is
underconstrained one can arrive at a multiplet which describes the
physical fields by imposing the further constraint that there should
exist appropriate $q$-form worldsurface gauge fields, $\cF_q$. We will
describe this constraint in the case of D$p$-branes below. 

In addition to the embedding matrix, each brane comes with a Wess-Zumino
form, $W_{p+2}$, defined on $M$. This term takes different forms for
different branes. To be specific, let us consider the fundamental
F$p$-branes and L$p$-branes \cite{hs1} with $16$ target space
supersymmetries and D$p$-branes and the M5-branes which have 32 target
space supersymmetries. The F-class corresponds to $p$-branes in $p+5$
dimensions $(p=1,2...,5)$ and the L-class corresponds to $p$-branes in
$p+4$ dimensions $(p=1,2,...,5)$. In each one of these cases there exist
Cartan integrable systems in the target space which take the form 

\be
\ba{lrclrclrcl}
Fp\ \ : \hspace{0.5cm}  &  d G_{p+2} &=& 0\ ,  & & & & & & 
\\[.3cm]
Lp\ \ : \hspace{0.5cm}  &  dG_{p+2} &=& G_2 G_{p+1}\ ,\hspace{0.5cm}
 & dG_{p+1} &=& 0\ , \hspace{0.5cm} dG_2 &=& 0\ , &
\\[.3cm]
Dp\ \ :  \hspace{0.5cm} &  dG &=& G H_3 \ , 
\hspace{1cm} &  dH_3 &=& 0 \ , &&&
\\[.3cm]
M5\ : \hspace{0.5cm} & dG_7 &=& G_4 G_4 \ , \hspace{1cm}
& dG_4 &=& 0 \ , &&&
\ea
\ee

where, in the D$p$-brane case, $G$ is a sum of the Ramond-Ramond (RR) curvatures which have
even/odd ranks in type IIA/B theory, and wedge products of forms are
understood. These equations can be solved locally to give

\be
\ba{lrclrclrcl}
Fp\ \ : \hspace{0.5cm}  &  G_{p+2} &=& dC_{p+1}\ ,  & & & & & & 
\\[.3cm] 
Lp\ \ : \hspace{0.5cm}  &  G_{p+2} &=& dC_{p+1}-C_1 G_{p+1} \ , &
G_{p+1} & = & dC_p\ ,        &  G_2 &=& dC_1\ , 
\\[.3cm]
Dp\ \ :  \hspace{0.5cm} &  G &=& dC - C H_3 + m e^{B_2} \ , 
\hspace{1cm} & H_3 &=& dB_2 \ ,  &&&
\\[.3cm]
M5\ : \hspace{0.5cm}  & G_7 &=& dC_6-C_3 G_4\ , 
\hspace{1cm} &  G_4 &=& dC_3 \ , &&&
\ea
\ee

where $m$ is an arbitrary constant which is relevant for type IIA theory
and $C$ is the sum of the RR potentials. We denote by $C$ the potentials
associated with all the target space field strengths, with the exception
of $H_3=dB_2$ which plays a special role in the case of D$p$-branes.

The Wess-Zumino form $W_{p+2}$ is a closed form

\be
dW_{p+2}=0\ ,  \la{wz}
\ee

constructed from from the pull-backs of suitable target space forms as
well as intrinsic worldvolume forms. For F$p$-branes the form $G_{p+2}$ is
closed, and therefore its pullback to the worldvolume is a candidate
Wess-Zumino form. However, the forms $G_{p+2}$ in D$p$-brane case,  $G_7$ in the M$5$-brane case and $G_{p+2}$ in the 
$L_p$-brane case  are not closed.
This is remedied by introducing respectively a two-form $\cF_2$, a three-form $\cF_3$ 
and a $p$-form $\cF_p$ as follows

\bea
\cF_2 &=& dA_1 - f^* B_2\ , \la{cf2}
\\
\cF_3 &=& dA_2 -f^* C_3\ , \la{cf3}
\\
\cF_p &=& dA_{p-1}- f^* C_p\ .\la{cfp}
\eea

These satisfy the Bianchi identities

\bea
d\cF_2 &=& - f^* H_3\ ,
\nn\\
d \cF_3 &=& -f^* G_4\ 
\nn\\
d \cF_p &=& -f^* G_{p+1}\ .
\eea

Note that the construction of these forms has led to
the introduction of intrinsic worlvolume potentials $A_1, A_2$ and
$A_{p-1}$. Using the ingredients described above, we construct the
Wess-Zumino forms as follows:

\be
W_{p+2}=\cases{f^* G_{p+2} &  \quad\quad F$p$ \cr
&\cr 
f^* \left(G_{p+2}\ + \cF_p G_2 \right) & \quad\quad L$p$ \cr
&\cr 
\left((f^* G) e^\cF \right)_{p+2} & \quad\quad   D$p$ \cr
&\cr
f^* \left(G_7 + \cF_3 G_4 \right) & \quad\quad M$5$ }
\ee

It is easy to verify that all these forms are indeed closed. Thus, the
Wess-Zumino form $W_{p+2}$ can {\it locally} be written as 

\be
W_{p+2} = d Z_{p+1}\ ,
\label{6}
\ee

where 

\be
Z_{p+1}=\cases{f^* C_{p+1} &  \quad\quad F$p$ \cr
&\cr 
f^* \left(C_{p+1} + C_1 \cF_p \right) & \quad\quad L$p$ \cr
&\cr   
 \left((f^* C) e^\cF \right)_{p+1} + m\o_{p+1} & \quad\quad  D$p$ \cr
&\cr
f^* \left( C_6 + C_3 \cF_3  \right) & \quad\quad M$5$}
\ee

and where $\o_{p+1}(A,dA)$ is the Chern-Simons form present for type IIA
D$p$-branes defined by

\be
d\o_{p+1}(A,dA)= (e^{dA})_{p+2}\ .
\ee

We mentioned earlier that for the cases where the worldsurface multiplet
is underconstrained one can arrive at a multiplet which describes the
physical fields by imposing the further constraint on a suitable
worldvolume superform. In the case of D$p$-branes that constraint is
\cite{chs}

\be 
\cF_{\a B}=0\ , \la{fc} 
\ee 

i.e. all of the components of $\cF$ except the purely bosonic ones must
vanish. It can be shown that for $p <6$ the basic embedding condition
puts the theory on-shell \cite{hrss}, and that for these cases the
condition \eq{fc} follows automatically. In other cases one can argue
for these constraints by considering open branes which end on other
branes \cite{chs,chsw}. A similar situation arises for the
M5-brane, for which we refer the reader to refs. \cite{hs2,cs}. The case
of L$p$-branes will be treated in detail elsewhere \cite{hrrs}.

%%%%%%%%%%%%%%%%%%%%%%%%%%%%%%%%%%%%%%%%%%%%%%%%%%%%%%%%%%%%%%%%%%%%%%%%%%%%

\section{Kappa Symmetry and Green-Schwarz Actions}

%%%%%%%%%%%%%%%%%%%%%%%%%%%%%%%%%%%%%%%%%%%%%%%%%%%%%%%%%%%%%%%%%%%%%%%%%%%%

The basic embedding condition \eq{2}, which underlies all branes studied so
far and which is geometrically very natural, is intimately related to
$\k$-symmetry in the GS approach to branes. Under an infinitesimal
worldsurface diffeomorphism one has

\be
(\d z^{\unM})E_{\unM}{}^{\unA}=v^A E_{A}{}^{\unA}
\label{3}\ ,
\ee

where $v^A$ is the worldsurface vector field generating the
diffeomorphism. For an odd diffeomorphism, with $v^a=0$, one finds,
using the embedding condition \eq{2},

\be
\d z^{\una}\equiv (\d z^{\unM})E_{\unM}{}^{\una}=0\ ,
\label{4}
\ee

and

\be
\d z^{\ua}\equiv (\d z^{\unM})E_{\unM}{}^{\ua}= v^{\a} E_{\a}{}^{\ua}\ .
\label{4a}
\ee

This can be rewritten in the more usual $\k$-symmetry  form  

\be
\d z^{\ua}= {1\over2}\k^{\ub}(1+\C)_{\ub}{}^{\ua}\ ,
\label{4b}
\ee

where

\be
\k^{\ua}=v^{\a} E_{\a}{}^{\ua}\ 
\label{4c}
\ee

and where

\be
P_{\ua}{}^{\ub}={1\over2}(1+\C)_{\ua}{}^{\ub}
\label{4d}
\ee

is the projection operator onto the odd tangent space of the
worldsurface from the odd tangent space of the target. It is given in
terms of $E_\a{}^{\ua}$ by

\be
P_{\ua}{}^{\ub}=(E^{-1})_{\ua}{}^{\c} E_{\c}{}^{\ub}
\label{4e}
\ee

Thus we have

\bea
\d z^{\una}&=&0\\
\d z^{\una}&=&{1\over2}\k^{\ub}(1+\C)_{\ub}{}^{\ua}
\label{4f}
\eea

Equations \eq{4f}, evaluated at $\th=0$, are the standard $\k$-symmetry
transformations of $z^{\unM}(x)$ in the GS formalism. The explicit form
of the operator $\C$, which must square to unity in order for $P$ to be
a projector, and the explicit relation between the parameters for
$\k$-symmetry and worldsurface supersymmetry depend on the choice of
basis for the odd tangent space on the worldsurface, but whichever basis
one chooses to work with, $\k$-symmetry will have a precise definition
in terms of worldsurface supersymmetry. Of course the latter does not
change and so should, we would argue, be thought of as being more
fundamental.

For any brane the Wess-Zumino form $W_{p+2}$ is closed. Since it is a
$p+2$-form on a manifold which has even (i.e. bosonic) dimension $p+1$
it follows that it is exact. This is so because the de Rham cohomology
of a supermanifold coincides with the de Rham cohomology of its body.
Therefore we can always write

\be
W_{p+2}= d K_{p+1}
\label{7}
\ee

for some {\it globally} defined $(p+1)$-form $K$ on $M$. Furthermore,
since none of the target space fields or the worldsurface fields has
negative dimension, at least for the models under discussion here, it
follows that the only non-vanishing component of $K$ is the purely
bosonic one. In components this means 

\be
K_{\a A_1 \cdots A_p}=0\ .
\ee

We now define the Green-Schwarz Lagrangian form $L_{p+1}$ to be

\be
L_{p+1}= K_{p+1}-Z_{p+1}
\label{8}
\ee 

Under a worldsurface diffeomorphism generated by the vector field $v$
one has

\be
\d L_{p+1}={\cal L}_v L_{p+1} =d i_v L_{p+1} + i_v d L_{p+1}
\label{9}
\ee

Since, by construction, $L_{p+1}$ is closed,

\be
dL_{p+1}=0\ ,
\ee

the variation \eq{9} reduces to

\be
\d L_{p+1}=d i_v L_{p+1}\ .
\la{9a}
\ee

Therefore the action integral

\be
S=\int_{M_0}\, L^0_{p+1}\ ,
\la{9b}
\ee

where $M_0$ is the body of $M$ and where 

\be
L^0_{p+1}=dx^{m_{p+1}}\wedge dx^{m_p} \wedge 
\ldots dx^{m_1} L_{m_1\ldots m_{p+1}}\vert\ ,
\la{9c}
\ee

where the vertical bars indicate evaluation of a (worldvolume)
superfield at $\th=0$, will be invariant under $\k$-symmetry
transformations and diffeomorphisms of $M_0$, since these
transformations are identified with the leading components of the
superdiffeomorphisms of $M$. 

As we noted in the introduction, this result is closely related to both
superspace actions of the heterotic string type and to the generalised
actions of refs \cite{a1,a4,a5}. However, there is a difference in that,
in the generalised action formalism \cite{a1,a4,a5}, the
Dirac-Born-Infeld action is explicitly included in the case of
$D$-branes. The method proposed here generates the DBI action (from
$K_{p+1}$) automatically, and moreover allows for the DBI action to be
extended to worldsurface $q$-form gauge fields with $q>2$. The argument
given above shows that the Lagrangian we have constructed is invariant
uder the right symmetries and has the usual Wess-Zumino term. The
contribution to the action from $K$ must therefore be the DBI action.
Below we shall show that this is indeed the case in specific examples.

It is worth emphasizing that not only is the DBI action automatically
generated in the method proposed here, but that also the $\k$-symmetry
of the total action is made manifest. This is due to the closure
property \eq{9a}. In the generalized action formalism, however, while
the action is indeed an integral of a Lagrangian $(p+1)$-form over
$M_0$, not only is the DBI term explicitly included (along with certain
Lagrange multiplier terms), but also the closure property $dL_{p+1}=0$,
needed for the proof of $k$-symmetry, is non-manifest, and proving it
requires lengthy calculations \cite{a4,a5}. 

The form of the Lagrangian given in \eq{8} is closely related to the
actions considered before in \cite{e7} for the heterotic string, in
\cite{e9} for the $D=11$ supermembrane, and in \cite{e10} for higher
super $p$-branes. We shall comment about this relation in more detail in
Section 7.

%%%%%%%%%%%%%%%%%%%%%%%%%%%%%%%%%%%%%%%%%%%%%%%%%%%%%%%%%%%%%%%%%%%%%%%%%%%%

\section{M2-brane}

%%%%%%%%%%%%%%%%%%%%%%%%%%%%%%%%%%%%%%%%%%%%%%%%%%%%%%%%%%%%%%%%%%%%%%%%%%%%

To illustrate the above general formalism we consider first the simplest
case, namely an on-shell type I brane, the membrane (M2-brane) in
$D=11$. We assume that the embedding condition \eq{2} holds. (See
\cite{bsm2} as well for a treatment of the supermembrane in the
superembedding formalism ). It can then be shown that we may choose

\bea
E_{\a}{}^{\ua} &=& u_{\a}{}^{\ua} \\
E_{a}{}^{\una}&=& u_a{}^{\una}
\label{9d}
\eea

with the complementary normal matrix $E_{A'}{}^{\unA}$, which specifies
the choice of normal spaces, being given by

\bea
E_{\a'}{}^{\ua} &=& u_{\a'}{}^{\ua} \\
E_{a'}{}^{\una}&=& u_a{}^{\una}
\label{9e}
\eea

We may also impose

\be
E_{\a'}{}^{\una}=0\ .
\la{9f}
\ee

In these formulae $u$ denotes an element of the group $Spin(1,10)$ or
the corresponding element of the Lorentz group in eleven dimensions.
Thus the matrices $u_{\a}{}^{\ua}$ and $u_{\a'}{}^{\ua}$ together make
up an element of $Spin(1,10)$ while $u_a{}^{\una}$ and $u_{a'}{}^{\una}$
make up the corresponding element of $SO(1,10)$. We remind the reader
that although $E_{\a}{}^{\una}=0$, it is not the case that
$E_{a}{}^{\ua}=0$, although we can choose

\be
E_{a}{}^{\ua}=\L_a{}^{\a'} u_{\a'}{}^{\ua}\ .
\la{9g}
\ee

The leading component of the superfield $\L_a{}^{\a'}$ should be thought
of as the spacetime derivative of the transverse fermionic
coordinate field, that is, the derivative of the physical fermion field
of the membrane. 

In order to derive the GS action from the superembedding formalism it is
necessary to show that the $\th=0$ component of $E_m{}^a$, which we
denote by ${\cE}_m{}^a$ is the dreibein for the GS metric. The latter is
defined to be

\be
g_{mn}=\left(\del_m z^{\unM} E_{\unM}{}^{\una}\right)
\left(\del_n z^{\unN} E_{\unN}{}^{\unb}\h_{\una\unb}\right) \vert\ ,
\la{9h}
\ee

where, we recall that the bar denotes evaluation of a quantity at
$\th=0$. From the embedding condition we have

\be
E_{\a}{}^{\una}=E_{\a}{}^m \left(\del_m z^{\unM} \right)E_{\unM}{}^{\una} 
+ E_{\a}{}^{\m}\left(\del_{\m} z^{\unM}\right) E_{\unM}{}^{\una}=0\ .
\la{9i}
\ee

We can always choose a gauge on the worldvolume such that
$E_{\a}{}^{m}|=0$. Moreover the leading component of $E_{\a}{}^{\m}$ is
non-singular. Therefore, evaluating the above equation at $\th=0$ we
deduce

\be
\del_{\m} z^{\unM} E_{\unM}{}^{\una}|=0\ .
\la{9j}
\ee

Using this result we find

\be
E_a{}^{\una}|= E_a{}^m \left(\del_m z^{\unM}\right) E_{\unM}{}^{\una}|
\ .
\la{9k}
\ee

It then follows, since $E_a{}^m|={\cE}_a{}^m$, the inverse of
${\cE}_m{}^a$, and the fact that

\be
E_a{}^{\una}E_b{}^{\unb}\h_{\una\unb}=\h_{ab}\ ,
\la{9l}
\ee

that ${\cE}_m{}^a$ is indeed the dreibein for the GS metric as claimed, i.e.

\be
{\cE}_m{}^a {\cE}_n{}^b \h_{ab}=g_{mn}\ .
\la{9m}
\ee

The Wess-Zumino form for the M2-brane is the pull-back of the
supergavity four-form $G_4$. Its non-vanishing components are

\be
G_{\ua\ub\unc\und}=-i(\C_{\unc\und})_{\ua\ub}
\la{9n}
\ee

and the totally vectorial component $G_{\una\unb\unc\und}$. On the
worldvolume of the brane there should therefore be a three-form $K_3$
such that

\be
W_4=f^* G_4= d K_3\ .
\la{9o}
\ee

In index notation this reads

\be
4\nab_{[A}K_{BCD]}+ 6T_{[AB}{}^E K_{|E|CD]}= (f^*G)_{ABCD}\ .
\la{9p}
\ee

This is indeed the case as we shall now verify. Since there are no
fields of negative dimension on the worldvolume (given the standard
embedding condition), the only non-vanishing component of $K$ has purely
vectorial indices. By directly evaluating the dimension zero component
of the above equation one finds that it is satisfied for

\be
K_{abc}=\e_{abc}\ .
\la{9q}
\ee

Since there are no fields of negative dimension it is apparent that the
negative dimension compoents of $W_4=dK_3$ are trivially satisfied. To
prove that the remaining components are also satisfied it is convenient
to introduce a four-form $I_4$ defined by 

\be
I_4=W_4- dK_3\ ,
\label{9r}
\ee

where $K_3$ has the components described above. Clearly $dI_4=0$. We
need to show that $I_4=0$ but, by dimensional analysis, the only
components of $I_4$ that need to be checked are $I_{\a\b cd}$ and $I_{\a
bcd}$ (since $I_{abcd}$ vanishes identically). The fact that $I_{\a\b
cd}$ vanishes can easily be checked using the formulae given above while
one can show that this implies automatically that $I_{\a bcd}=0$ by
using the identity $dI_4=0$. In a coordinate basis one therefore has 

\be
K_{mnp}|=\e_{mnp} \sqrt{-\det g}\ , 
\la{9s} 
\ee 

where $g$ is the GS metric. The GS Lagrangian is therefore recovered
from the general formulae \eq{8} and \eq{9b}; it is 

\be {\cL}=\sqrt{-\det g} -{1\over 6}\e^{mnp}\del_p
z^{\unP}\ \del_n z^{\unN}\ \del_m z^{\unM}\ C_{\unM\unN\unP}\ , 
\la{9t} 
\ee

where $G_4=dC_3$ on $\unM$, and where 

\be 
L^0=dx^m\wedge dx^n\wedge dx^p \e_{mnp}\ {\cL}\ .
\la{9u} 
\ee

%%%%%%%%%%%%%%%%%%%%%%%%%%%%%%%%%%%%%%%%%%%%%%%%%%%%%%%%%%%%%%%%%%%%%%%%%%%%

\section{D2-brane}

%%%%%%%%%%%%%%%%%%%%%%%%%%%%%%%%%%%%%%%%%%%%%%%%%%%%%%%%%%%%%%%%%%%%%%%%%%%%

The on-shell example we shall consider is the IIA $D2$-brane in $D=10$.
For simplicity we shall take the target space to be flat and $m=0$,
although this is not essential. The basic embedding equation \eq{2} is
imposed as usual and we may choose to parametrise the dimension zero
components of the embedding matrix in the form \cite{hs2}

\bea
E_{\a}{}^{\ua} &=& u_{\a}{}^{\ua} + h_{\a}{}^{\b'} u_{\b'}{}^{\ua} \nn\\
E_{a}{}^{\una}&=& u_a{}^{\una}\ .
\label{10}
\eea

Here $u$ denotes part of a matrix of the group $Spin(1,9)$, in the
spinor or the vector representations according to the indices. The
pull-back to the worldsurface of the defining equation for the target
space torsion two-form gives the equation

\be
\nab_A E_{B}{}^{\unC}-(-1)^{AB}\nab_B E_{A}{}^{\unC} 
+ T_{AB}{}^C E_C{}^{\unC}=(-1)^{A(B+\unB)} E_B{}^{\unB} E_A{}^{\unA} 
T_{\unA\unB}{}^{\unC}\ .
\label{11}
\ee

The dimension zero component of this equation reads, on using the
embedding condition \eq{2},

\be
E_{\a}{}^{\ua}E_{\b}{}^{\ub} T_{\ua\ub}{}^{\unc}=T_{\a\b}{}^c
E_c{}^{\unc}\ .
\label{12}
\ee

Using

\be
T_{\ua\ub}{}^{\unc}= -i(\C^{\unc})_{\ua\ub}
\label{13}
\ee

and \eq{10} one finds \cite{hrss}

\be
h_{\a}{}^{\b'}\rightarrow h_{\a i}{}^{\b' j}
= \d_i{}^j(\c^{ab})_{\a}{}^{\b'} h_{ab}
\label{14}
\ee

and

\be
T_{\a\b}{}^c=-i(\C^d)_{\a\b} m_d{}^c\ ,
\label{15}
\ee

where

\be
m_a{}^b=\d_a{}^b(1-4y) + 8(h^2)_a{}^b
\label{16}
\ee

with

\be
y={1\over 2} {\rm tr}\, h^2
\label{17}
\ee

and where $h^2$ denotes matrix multiplication. 

It is not difficult to show that the embedding condition \eq{2} implies the
existence of a two-form $\cF$ such that

\be
d\cF=-f^*H_3\ ,
\label{18}
\ee

where $H_3$ is the pull-back of the NS three-form on the target space.
This identity is satisfied provided that we choose all the components of
$\cF$ to vanish except for $F_{ab}$ which is related to $h$ by

\be
m_a{}^c \cF_{cb}= 4h_{ab}\ .
\label{19}
\ee

This can be rearranged to give

\be
\cF_{ab}={4 h_{ab}\over 1+4y}\ . 
\la{fh}
\ee

The Wess-Zumino four-form $W_4$ is given by

\be
W_4=d Z_3= d(f^*C_3 + f^*C_1 \cF)\ ,
\label{20}
\ee

where $C_1$ and $C_2$ are two of the RR potentials on the target space.
It can be rewritten as

\be
W_4 = f^*G_4 +f^*G_2 \cF\ ,
\label{21}
\ee

where the RR field strengths $G_4$ and $G_2$ are given by

\bea
G_4&=& d C_3 + B_2 G_2 \\
G_2&=& d C_1
\label{22}
\eea

with $B_2$ being the potential for the NS field strength $H_3$. The
non-vanishing components of the RR fields in flat superspace are

\bea
G_{\ua\ub\unc\und}&=& -i(\C_{\unc\und})_{\ua\ub} \\
G_{\ua\ub}&=&-i(\C_{11})_{\ua\ub} 
\label{23}
\eea

It is now straightforward to verify that 

\be
W_4= d K_3
\label{24}
\ee

where all of the components of $K_3$ vanish except for $K_{abc}$ which
is given by

\be
K_{abc}=\e_{abc}\ K\ ,
\ee

with

\be
K={1-4y\over 1+4y}\ .
\label{25}
\ee

The part of the GS Lagrangian arising from $K$ is then given by

\be
K_{mnp}|= (E_m{}^a E_n{}^b E_p{}^c )\e_{abc}\ K|\ .
\label{27}
\ee

However, $E_m{}^a|={\cE}_m{}^a$ is again simply the dreibein for the
induced GS metric,

\be
{\cE}_m{}^a{\cE}_n{}^b\h_{ab}=g_{mn}\ .
\label{28}
\ee

From this we derive

\be
K_{mnp}|=\e_{mnp}\, \sqrt{ -{\rm det}\,g}\, \times K|\ .
\label{29}
\ee

It remains to show that $K$ is proportional to $\det\
\sqrt{\d_m{}^n+{\cF}_{m}{}^{n}}$. To this end, we first observe that we
can replace the coordinate indices appearing in the Born-Infeld
determinant as written here by orthonormal ones at no cost. Thus we can
work with $\cF_{ab}$ and $h_{ab}$ and then evaluate at $\th=0$. We have

\be
\ba{rcl}
\det (1 + \cF )&=&\exp {\rm tr} \log \left(1 +\cF \right) \\[.3cm]
               &=&\exp {\rm tr} \left( -{\cF^2\over2} 
	+ {\cF^4\over4}-\dots\right)\ ,
\ea
\la{30a}
\ee

where the second step follows from the antisymmetry of $\cF$. Writing
$\cF$ in terms of $h$ using \eq{fh} and employing the identity

\be
h^3= y h\ ,
\la{30b}
\ee

we find

\be
\ba{rcl}
\det (1 + \cF )&=&\exp {\rm tr}\, 
h^2 \left({-4^2\over2(1+4y)^2} +{4^4\over4(1+4y)^4}-\ldots \right)\\
&&\\
&=&\exp \log \left(1-{16y\over (1+4y)^2} \right)\\
&&\\
&=& \left({1-4y\over 1+4y}\right)^2\ .
\ea
\la{30c}
\ee

Therefore, $K$ defined in \eq{25} is given by

\be
K=\sqrt{{\rm det}(\d_m{}^n + \cF_m{}^n)}\ ,
\label{30}
\ee

and that the GS Lagrangian is obtained from the general formulae \eq{8}
and \eq{9b} to be

\be 
L^0=dx^m\wedge dx^n\wedge dx^p \e_{mnp}\ {\cL}\ .
\la{9uu} 
\ee
with

\be
{\cL}= \sqrt{-{\rm det}(g_{mn} + \cF_{mn})} -f^*C_3 - f^*C_1 \cF\ ,
\ee

in agreement with the general results for D$p$-brane actions in the GS
formalism \cite{d0,d1,d2,d3,d4}.

%%%%%%%%%%%%%%%%%%%%%%%%%%%%%%%%%%%%%%%%%%%%%%%%%%%%%%%%%%%%%%%%%%%%%%%%%%%%

\section{The Membrane in $N=1,\ D=4$ Superspace}

%%%%%%%%%%%%%%%%%%%%%%%%%%%%%%%%%%%%%%%%%%%%%%%%%%%%%%%%%%%%%%%%%%%%%%%%%%%%

The final example we shall consider is the membrane in $N=1,\ D=4$
superspace. This is a type I brane for which the standard embedding
condition defines an off-shell multiplet. Actually, this brane has
codimension one and the worldsurface multiplet in question is an entire
scalar superfield, but this is is simply the off-shell scalar multiplet
in three dimensions. 

To simplify the discussion we shall take the target space to be flat.
For the embedding matrix we can take, as before,

\bea
E_{\a}{}^{\una}&=&0 \\
E_{\a}{}^{\ua} &=& u_{\a}{}^{\ua} + h_{\a}{}^{\b'} u_{\b'}{}^{\ua}\ , 
\label{33}
\eea

where both $\a$ and $\a'$ are $d=3$ spinor indices taking two values. We
also choose

\be
E_a{}^{\una}=u_a{}^{\una}\ .
\la{33a}
\ee 

The dimension zero component of the torsion equation \eq{11} gives

\be
h_{\a}{}^{\b'}=i\d_{\a}{}^{\b'}h
\label{34}
\ee

and

\be
T_{\a\b}{}^c=-i(1+h^2)(\c^c)_{\a\b}\ ,
\label{34a}
\ee

but in this case, the (real) field $h$ is not related to a gauge field,
rather its leading component is the auxiliary field in the scalar
multiplet. 

The Wess-Zumino form $W_4$ is in this case simply the pull-back of the
target space four-form $G_4=dC_3$ to $M$. The only non-vanishing
component of $G_4$ for a flat target space is

\be
G_{\ua\ub \unc\und}=-i(\C_{\unc\und})_{\ua\ub}\ .
\label{35}
\ee

The general argument given previously implies that

\be
W_4=d K_3\ .
\label{36}
\ee

It is straightforward to verify that this is indeed the case, and that
the only non-vanishing component of $K_3$ is

\be
K_{abc}=\e_{abc}\ K\ ,
\label{37}
\ee

where

\be
K={1-h^2\over 1+h^2} \ .
\label{37a}
\ee

The GS Lagrangian form is

\be
L_{mnp}=(K_{mnp} - (f^*C)_{mnp})|\ ,
\label{38}
\ee

from which the GS Lagrangian density is found to be

\be
{\cL}=\sqrt{-\det g}\ \left({1-h^2\over 1+h^2} \right)\vert
-{1\over6}\e^{mnp}(f^*C)_{mnp}\vert\ ,
\label{39}
\ee

where $g$ is again the standard GS induced metric. The only difference
from the usual GS Lagrangian is the factor multiplying the GS measure
containing as it does the auxiliary field $h$. However, the equation of
motion for this field is purely algebraic and can be used to set $h=0$.
We thus recover the standard GS action. It is amusing to note that the
off-shell action given here has the same form as the DBI action for the
D2-brane when expressed in terms of $h$. 

Since the multiplet is off-shell it is possible to construct a
superspace action for this model using the techniques that were
introduced in \cite{e7} in the context of the heterotic string. In order
to do this it is useful to introduce the notion of a $q$-vector density,
which we shall call a $q$-coform for short. Such an object is a tensor
density of tensorial type (q,0) which is totally antisymmetric. There is
a natural pairing between $q$-coforms $P$ and $q$-forms $\o$ given by

\be
(P,\o) =\int P^{M_q\ldots M_1}\o_{M_1\ldots M_q}\ .
\label{40}
\ee

If the space of $q$-coforms is denoted by $\tilde\O_q$, then there is a
natural derivative $\tilde{d}$ which maps $\tilde\O_q$ to
$\tilde\O_{q-1}$ and which satisfies $\tilde{d}^2=0$. In a coordinate
basis one has

\be
(\tilde{d}P)^{M_1\ldots M_{q-1}}=P^{M_1\ldots M_q},_{M_q}
:=P^{M_1\ldots M_q} {\stackrel{\leftarrow}{\del}}_{M_q}\ ,
\label{41}
\ee

where the derivative is the right derivative. One then has

\be
(P,d\o) = -(\tilde{d}P,\o)\ ,
\label{42}
\ee

up to possible surface terms. Using this notation we can write an action
for the membrane in the form

\be
S= (P, K-f^*C -dQ)\ ,
\label{43}
\ee

where $P$ is a three-coform and $Q$ a new two-form field. This action is
invariant under

\be
P\rightarrow P + \tilde{d} X
\label{44}
\ee

where $X$ is a four-coform. Varying the action with respect to $Q$ gives

\be 
\tilde{d}P=0
\label{44a}
\ee

Thus $P$ is an element of the third homology group associated with the
operator $\tilde{d}$. This group is one-dimensional, but non-trivial,
and can be represented in suitable coordinates by

\be
P^{mnp}=\th^2 \e^{mnp} \times {\rm constant} \la{p3}
\ee

with all other components vanishing. The constant of integration is
naturally interpreted as the tension of the brane. Varying the action
with respect to the embedding is quite complicated, but because we know
the content of the worldsurface multiplet we can instead substitute
\eq{p3} back into the action to get the Green-Schwarz form given above.

In constructing this action we have assumed that the basic embedding
condition \eq{2} is satisfied. However, one can also derive this by
including a Lagrange multiplier field $\P_{\una}{}^{\a}$ to impose the
embedding constraint. The action then becomes

\be
S=\int \Pi_{\una}{}^{\a} E_{\a}{}^{\una} -(P, K-f^*C-dQ)
\label{46}
\ee

This action has precisely the same structure as the action given for the
heterotic string. Although such actions have been written down
previously for $p$-branes with $p\geq2$, it was not known at the time
whether such actions would lead to the corerct brane dynamics. If the
embedding condition imposed by the Lagrange multiplier superfield leads
to an on-shell world-volume multiplet then more degrees of freedom,
contained within the Lagrange multiplier superfield, will also
propagate. The example studied here is an off-shell multiplet and so
would seem to provide the first well-established example of a superfield
action for a brane with $p>1$.

We conclude this section with two comments. Firstly, the form of the
action integral given in \eq{43} can also be used when the constraints
are on-shell. In such a case, the resulting action should not be thought
of as a superfield action, but simply as the GS action rewritten in
superspace. Secondly, in order to write down the action for the $D=4$
membrane in the form given in \eq{46}, the embedding constraint must be
relaxed. However, even in this case there will still be a globally
defined three-form $K$ such that $dK_3=W_4$, although the explicit form
for $K$ will be more complicated than it is when the embedding condition
is satisfied.

\bigskip

%%%%%%%%%%%%%%%%%%%%%%%%%%%%%%%%%%%%%%%%%%%%%%%%%%%%%%%%%%%%%%%%%%%%%%%%%%%%

\section{Comments}

%%%%%%%%%%%%%%%%%%%%%%%%%%%%%%%%%%%%%%%%%%%%%%%%%%%%%%%%%%%%%%%%%%%%%%%%%%%%

In this paper we have shown that one can construct Green-Schwarz actions
for almost all branes, excluding those with self-dual gauge fields, in a
systematic fashion starting from the Wess-Zumino form on the
worldvolume. It is important that one uses the superembedding formalism
to derive this result because the Wess-Zumino form vanishes identically
on the bosonic worldvolume. The resulting superspace Lagrangian form,
$L_{p+1}$, given in \eq{8} can be obtained explicitly if one know the
$(p+1)$-form $K_{p+1}$ for a given brane. In fact, if the standard
embedding condition \eq{2} holds, it is straightforward to invert the
relation 

\be
W_{p+2}=d K_{p+1}
\label{46a}
\ee

This is because the only non-vanishing component of $K$ in this case
will be the one which has only vectorial indices while the only
non-vanishing component of $W$ will be the one with two spinor indices
and $p$ vectorial indices, provided that the background geometry is of
standard type (which would be expected to arise form brane integrability
in any case). In this situation one would have

\be
W_{\a\b c_1\ldots c_p}= T_{\a\b}{}^{c_o} K_{c_o c_1\ldots c_p}
\label{46b}
\ee

From this equation one determines $K$ to be

\be
K_{a_1\ldots a_{p+1}}=t_{a_1}{}^{\a\b} W_{\a\b a_2\ldots a_{p+1}}
\label{46c}
\ee

where $t_{a}{}^{\a\b}$ is the inverse of $T_{\a\b}{}^c$,

\be
t_{a}{}^{\a\b}T_{\a\b}{}^b=\d_a{}^b .
\label{46d}
\ee

Note that the right-hand side of this equation is totally antisymmetric
on the vector indices although this is not manifest.

This explicit form for $K$ was found previously in the case of
F$p$-branes in \cite{e7,e9,e10} where the Lagrangian form was referred
to as $\tilde{B}$. For these branes the dimension zero worldvolume
torsion and the inverse tensor $t$ have components proportional to the
components of the Dirac gamma matrices. In all these case one then finds

\be
K_{a_1\ldots a_p}=\e_{a_1\ldots a_p}
\label{46e}
\ee

and this in turn results in the standard Nambu kinetic term in the GS
action proportional to the bosonic worldvolume in the induced metric. 

In the general case $T_{\a\b}{}^c$ involves the worldvolume field $h$
and so the expression for $K$ will be more complicated. It will be equal
to the epsilon tensor times a scalar factor which, for example in the
case of D-branes, will be the Born-Infeld function as we saw explicitly
in the case of the D$2$-brane. However, it is worth emphasizing that in
our formulation no knowledge of the DBI action is assumed and it is
derived from the first principles described in the paper. Consequently
our formalism can be applied to construct new brane actions which will
involve generalisations of DBI actions with higher rank field strengths.
The formulation of \cite{a4}, applied to D$p$-branes, in essence
provides the form $K_{p+1}$ as the Dirac-Born-Infeld (DBI) kinetic term.
However, it should be emphasized that the knowledge of the usual GS type
formulation of the D$p$-brane action is used as an input in this
construction, thereby essentially elevating the known DBI action to an
integral in a bosonic slice of the worldvolume supermanifold. 

We have also shown that if the embedding condition leads to an off-shell
multiplet one can construct superspace actions using the ideas
introduced in \cite{e7} and illustrated this with the example of the
membrane in $D=4$. It would be interesting to determine the gauge fixed
action for this membrane in terms of worldvolume superfields in the
static gauge. This should give a manifestly supersymmetric superspace
action expressed in terms of physical superfields. One would hope to
recover in this way a manifestly supersymmetric, superfield formulation
of the results obtained some time ago in \cite{pktm2} by gauge fixing
the Green-Schwarz action accompanied by a complicated set of field
redefinitions.

The calculation of $K_7$ for the M$5$-brane would also of considerable
interest, because a self-dual worldvolume 3-form field strength is
involved. Manifestly Lorentz invariant actions for such forms do not
exist unless one introduces an auxiliary scalar field and new gauge
invariances, fixing of which necessarily breaks Lorentz invariance
\cite{ma1,ma2}. A generalized action principle fails due to the very
presence of this auxiliary field \cite{volkov}. On the other hand, a
generalized action principle for self-dual supergravity in
six-dimensions is known to exist \cite{df6}. The price one pays is that
the action is not supersymmetric when restricted to $x$-space. Instead,
one should vary the action first in the full group manifold, and then
restrict the result to $x$-space and this leads to supersymmetric and
consistent equations of motion. Determining $K_7$ for the M$5$-brane
would shed light on the question of whether a similar phenomenon could
occur in the M5-brane. In this context, it is worth mentioning the work
of \cite{sm5} where $\k$-symmetric M$5$-equations of motion are obtained
from an action which is not $\k$-invariant as it contains three-forms of
both dualities. 

To conclude, we emphasize that the action formula proposed in this paper
is a universal one which applies to all branes with the possible
exception of those which contain chiral forms. In addition to producing
the known brane actions in the GS formalism (upon restriction to bosonic
wolrdvolume), our action fomula solves the problem of constructing
actions for branes whose worldvolume supports supermultiplets that
contain higher than second rank antisymetric tensors. Indeed, we shall
apply the action formula of this paper to construct an action for an
L$5$-brane in $D=9$ which contains a linear multiplet with a four-form
potential \cite{hrrs}. Other applications of the action formula might
yield new insights into duality symmetries that map branes into other
branes, as well as facilitating the study of interesting residual
symmetries such as superconformal symmetry in various backgrounds of
interest.

\bigskip

\noindent{\bf Acknowledments}

We thank Dmitri Sorokin for useful discussions.

\pagebreak

\end{document}